\documentclass[10pt,conference,letterpaper]{IEEEtran}
\IEEEoverridecommandlockouts
\usepackage{cite}
\usepackage{amsmath,amssymb,amsfonts}
\usepackage{algorithm}
\usepackage{algpseudocode}
\usepackage{graphicx}
\usepackage{textcomp}
\usepackage{xcolor}
\usepackage{url}
\def\BibTeX{{\rm B\kern-.05em{\sc i\kern-.025em b}\kern-.08em
    T\kern-.1667em\lower.7ex\hbox{E}\kern-.125emX}}
\usepackage{listings}
\usepackage{xcolor}

\lstdefinestyle{yaml}{
  basicstyle=\ttfamily\scriptsize,
  breaklines=true,
  frame=single,
  columns=fullflexible,
  keepspaces=true,
  showstringspaces=false,
  keywordstyle=\color{blue},
  commentstyle=\color{gray},
  stringstyle=\color{orange}
}
\begin{document}

\makeatletter
\def\ps@IEEEtitlepagestyle{%
  \def\@oddfoot{}%
  \def\@evenfoot{}%
  \def\@oddhead{%
    \parbox[b]{\textwidth}{\centering
    Author copy of paper published at \textit{34th International Symposium on the Modeling, Analysis, and Simulation of Computer and Telecommunication System (MASCOTS2026)}}
  }%
  \def\@evenhead{\@oddhead}%
}
\makeatother

\title{
InFactPlanner: Planning Sustainable Geo-Distributed LLM Data Centers
    \vspace*{-.2\baselineskip}
}

\author{
\IEEEauthorblockN{\parbox{\linewidth}{\centering 
Nicoletta Tsiopani, Moysis Symeonides,  
George Pallis,
Marios D. Dikaiakos}}
\IEEEauthorblockA{
\IEEEauthorrefmark{1} Department of Computer Science\\
University of Cyprus\\
Email: \{ntsiop01, msymeo03, pallis, mdd\}@ucy.ac.cy}
    \vspace*{-2.2\baselineskip}
}

\maketitle

\begin{abstract}
The rapid growth of LLM inference is shifting sustainability concerns from one-time training to continuous serving, where infrastructure 
decisions shape energy use, carbon emissions, water consumption, and service quality. Yet operators often need to compare deployment alternatives before large-scale infrastructure is built, making direct measurement costly, slow, and sometimes infeasible. We present InFactPlanner, a trace-driven decision-support framework for what-if analysis of sustainable AI data center deployment
for LLM inference across single and geo-distributed sites. InFactPlanner combines query traces, hardware–model profiles, candidate site configurations, PUE/WUE parameters, renewable generation models, and time-varying grid carbon intensity to estimate power, energy, carbon emissions, water use, latency, and server utilization. The framework abstracts low-level serving effects into configurable hardware-model profiles, enabling rapid comparison of site selection, capacity placement, hardware, model, renewable integration, and routing choices. We validate the energy accounting pipeline by reproducing reference LLM inference energy estimates with less than 10\% deviation, evaluate scalability across multiple data centers and server counts, and demonstrate scenario-driven decision analyses for hardware selection, renewable placement, geographic deployment, and 
carbon-aware routing. Our results show that sustainability-optimal choices can differ from latency-optimal ones, and that the carbon value of deployment
 depends strongly on the local grid mix.\end{abstract}


    \vspace*{-.3\baselineskip}
\section{Introduction}
    \vspace*{-.3\baselineskip}

The rapid expansion of generative AI has triggered an unprecedented rush to develop new hyperscale data centers~(DCs). These facilities (often termed AI Factories) are expected to provide the computational capacity required to train increasingly large and capable large language models~(LLMs), as well as to support the rapidly growing inference workloads generated by users and LLM-powered applications~\cite{Stojkovic2025}. However, energy availability is emerging as a critical constraint on the pace of DC development. DCs already consume a significant amount of electricity, estimated at 415 TWh in 2024 or approximately 1.5\% of global electricity need~\cite{goldman2023, scientificamerican2023}.  This share is projected to more than double to around 945 TWh by 2030, reaching just under 3\% of global electricity consumption~\cite{iea2025:energyai}. As new AI-oriented DCs are deployed, their increasing demand for large, reliable, and continuous power supply is placing growing pressure on electricity grids, particularly in terms of generation capacity, transmission infrastructure, and grid interconnection. 

However, with generative AI evolving from LLM-based chatbots toward agentic applications in areas such as robotics, manufacturing, and ambient AI, workloads will increasingly require low latency, local context awareness, and real-time interaction with sensors and actuators. For such workloads, exclusive reliance on remote hyperscale DCs will often be impractical. Future AI infrastructure is therefore likely to combine centralized hyperscale facilities running a mixture of training and inference workloads with a dense fabric of small and medium-sized AI data centers supporting inference, agent coordination, and real-time decision-making at the edge. This expansion changes not only where AI computation takes place, but also how its resource demands accumulate across the electricity, cooling, and water systems that support it. 

Also, as AI infrastructure scales up, its environmental footprint becomes a growing concern. For example, the 2026 AI Index Report estimates that training a single frontier model such as Grok 4 may generate approximately 72k tons of $CO_2$ emissions, comparable to the annual emissions of about 17k cars\cite{aiindex2026}. 
Inference workloads are also causing environmental concern. While inference workloads are far less energy-intensive per run than frontier-model training, they run continuously and 
at a global scale~\cite{psu2023}. Therefore, they are expected to dominate AI DCs’ operational footprint and drive energy, carbon, and water use. Indeed, recent findings show that LLM inference accounts for over half of LLM life-cycle carbon emissions~\cite{ozcan2025}, other estimates suggest that GPT-4o-scale inference could consume more water annually than the drinking needs of 12 million people~\cite{jegham2025}. Consequently, sustainability has become a critical design problem for highly complex LLM serving infrastructures, where hardware selection, model choice, workload placement, cooling strategy, renewable-energy availability, and grid carbon intensity jointly determine environmental impact and service quality. A key mitigation strategy is to combine energy-efficient hardware and cooling with the use of renewable energy sources~(RES) and geo-distributed deployment. Geo-distribution is particularly important because RES availability is location- and time-dependent. 
So, distributed serving across sites lets
operators place or shift workloads toward locations with cleaner electricity mixes, higher renewable availability, lower cooling demand, or acceptable latency characteristics.
However, evaluating these trade-offs before deployment is difficult. 
Direct measurement requires representative infrastructure, realistic workloads, and site-specific data on electricity mix, renewable sources, carbon intensity, cooling conditions, and latency. Such requirements are often infeasible during early-stage planning, or otherwise costly and time-consuming to satisfy. Simulation and analytical modeling offer a practical way to compare deployment options and sustainability–performance trade-offs before infrastructure decisions~\cite{luo2013}.


Several recent studies examine the energy consumption and environmental footprint of LLM inference and data-center operation. These works provide important insights into specific parts of the overall deployment problem but largely remain siloed. Therefore,  substantial effort is required to combine them into a unified analysis 
framework at a common time resolution. To the best of our knowledge, no prior system offers an end-to-end framework for evaluating these factors jointly before deployment. 
InFactPlanner addresses this gap by introducing an abstraction layer for deployment-level reasoning, enabling operators to compare infrastructure configurations and assess sustainability–performance trade-offs before committing to deployment decisions.
The contributions of this work are threefold. First, we introduce an infrastructure-level sustainability analysis model for LLM inference that connects workload properties, hardware-model serving profiles, facility parameters, renewable generation, grid carbon intensity, and water use at a common time resolution. Second, we implement this model in InFactPlanner, a configurable and modular trace-driven framework that supports single-site and geo-distributed what-if analysis across alternative siting, hardware, model, renewable, and routing decisions. Third, we evaluate the framework through accounting validation, scalability analysis, and decision-oriented case studies, showing when latency-, energy-, and carbon-optimal deployment choices diverge. 

\vspace*{-.1\baselineskip}
\section{System Model \& Motivation}
\vspace*{-.1\baselineskip}

\begin{figure}[!t]
    \centering
    \includegraphics[width=0.85\linewidth]{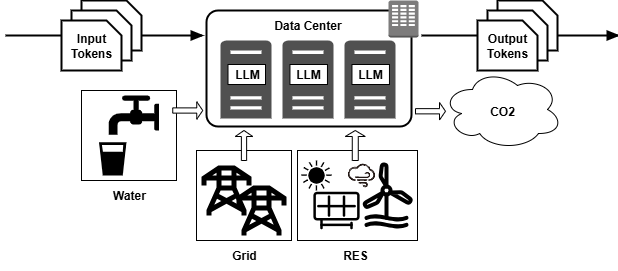}
    \vspace*{-1.\baselineskip}
    \caption{A DC as a system.}
    \label{fig:datacenter}
    \vspace*{-1.3\baselineskip}
\end{figure}

At a high level, a DC serving LLM-inference requests can be modeled as a deployment site that receives a timestamped stream of tokens carrying the requests, and serves them using GPU-accelerated servers (Fig.~\ref{fig:datacenter}). Each request is characterized by its arrival time, number of input tokens, and number of generated tokens. The DC maps these requests to one or more server groups, where each group is defined by a hardware type, an installed LLM, a concurrency limit, and a replication factor. The profile of each request determines the IT power demand induced at a given site; this demand is then used to estimate total facility energy consumption. The associated environmental footprint depends on the energy sources providing electricity to the site. A DC may draw energy directly from the grid, from local RES, or from a combination thereof. When local RES generation is insufficient, the residual demand is supplied by the grid. The corresponding carbon emissions depend on the time-varying carbon intensity of the local electricity mix, while water consumption depends on both on-site cooling requirements and off-site water use associated with electricity generation. Thus, the same LLM workload may lead to different environmental outcomes depending on the location, hardware, model, and renewable capacity.

This setting results in a complex infrastructure-planning problem for operators who seek to decide where to place LLM serving capacity across candidate data-center sites. An operator designing an LLM inference deployment must evaluate whether to serve demand from a single site or from multiple geo-distributed sites, which GPU servers and LLMs to deploy, whether and how to integrate renewable energy sources (RES), and how to route incoming requests. These choices affect latency, utilization, energy consumption, carbon emissions, and water use in different ways. For example, a high-performance GPU may reduce latency but increase total power demand, while deploying RES in a carbon-intensive region may reduce emissions more than deploying the same RES capacity in a region with a cleaner grid.

Direct evaluation of these alternatives through physical deployments is impractical during early-stage planning, as it would require representative infrastructure, realistic query traces, site-specific weather and grid data, and measurements across many combinations of models, hardware platforms, and locations. Manually combining existing analytical models is also difficult: LLM inference energy models, RES generation models, grid carbon-intensity, and water-use models are typically created in isolation and must be integrated at a common time resolution before they can support what-if analysis.

InFactPlanner addresses this gap through a deployment-level analytical framework for comparative what-if analysis of LLM inference deployments. Rather than modeling low-level GPU execution in detail, it abstracts serving behavior through a configurable library of hardware-model profiles that capture token processing throughput, first-token latency, consumed power, and concurrency. These profiles are combined with inference traces, site configurations, RES models, grid carbon-intensity data, and PUE/WUE parameters to estimate latency, power, energy consumption, carbon emissions, and water use. The framework is therefore suited to early-stage infrastructure planning, where the goal is to compare deployment choices consistently rather than reproduce every detail of a production serving stack.
This design prioritizes scalable comparison over exact prediction of production metrics. To this end, InFactPlanner abstracts away fine-grained serving effects, including batching dynamics, KV-cache behavior, and prefill/decode phase separation. These abstractions may introduce deviations from real deployments, which we quantify in Section~\ref{sec:eval}.

\section{Related Work}

Several recent works study LLM inference energy and environmental impact across several dimensions.
Energy-focused studies benchmark LLM inference across models and GPUs identifying parameters that affect efficiency~\cite{samsi2023}, while broader footprint analyses consider energy, carbon, and water use~\cite{Peng2023,jegham2025}. However, they do not provide configurable deployment-level analysis across geo-distributed sites, RES integration, carbon, water, and QoS trade-offs.
Mitigation and simulation tools address complementary parts of this problem. SHIELD~\cite{Qi2024} co-optimizes carbon emissions, water use, and energy consumption in geo-distributed DCs, but does not focus on LLM inference. 
Moreover, DynamoLLM~\cite{Stojkovic2025} optimizes LLM inference clusters for energy efficiency under latency SLOs, but does not model geo-distributed deployments, renewable integration, or water needs. Ozcan et al.~\cite{ozcan2025} combine LLM inference and carbon-aware simulation to estimate energy consumption and carbon emissions, while Splitwise~\cite{Patel2024} simulates phase-split LLM inference with a focus on latency and throughput. 
LLMServingSim 2.0~\cite{cho2026llmservingsim} also supports heterogeneous and disaggregated LLM-serving deployments, modeling batching, routing, memory management, and power consumption. However, these detailed execution-oriented simulators can incur longer simulation times, making them less suitable for rapid early-stage deployment planning. 
While prior works study either LLM serving behavior or sustainability-aware DC operation separately, InFactPlanner unifies these perspectives in a deployment-level framework for trace-driven analysis of performance, geo-distribution, renewable integration, and environmental impact. In this context, we do not provide direct quantitative comparisons against execution-oriented simulators such as Splitwise or LLMServingSim because these systems target different abstraction layers and optimization objectives. 
Instead, we position InFactPlanner as a complementary framework that prioritizes rapid what-if analysis and large-scale deployment exploration over low-level serving fidelity.









\begin{figure}[!t]
    \centering
    \includegraphics[width=0.8\linewidth,
    trim={0.03cm 0.04cm .01cm 0.1cm},
    clip]{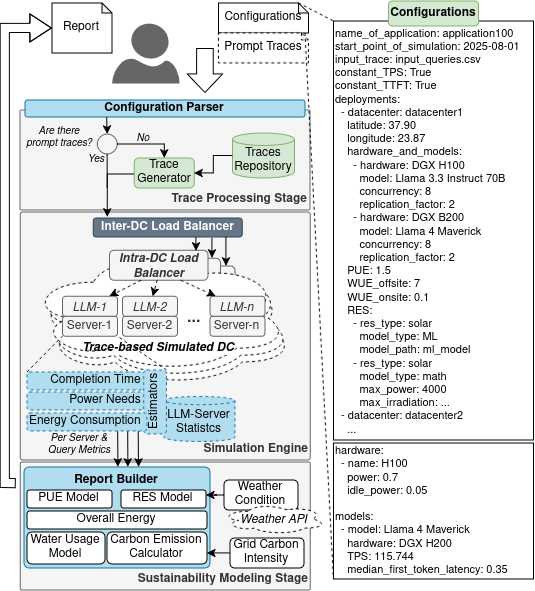}
    \vspace*{-.7\baselineskip}
    \caption{InFactPlanner Framework Overview.}
    \label{fig:framework}
    \vspace*{-1.5\baselineskip}
\end{figure}

\vspace*{-.1\baselineskip}
\section{The Framework Overview}
\vspace*{-.1\baselineskip}

InFactPlanner ("Inference (AI) Factory Planner") is organized as a trace-driven pipeline that transforms deployment descriptions and LLM inference workloads into sustainability and performance reports. Its 
workflow and main components are shown in Fig.~\ref{fig:framework}. InFactPlanner receives two main inputs: (i) a deployment configuration encoded in YAML, describing one or more DCs, and (ii) an input trace describing the expected LLM inference workload over time. The \textit{Configuration Parser} validates the DC deployment configuration and instantiates the DC deployment model. 

The \textit{Trace Processing Stage} prepares the workload to be used by InFactPlanner's Simulation Engine. 
When detailed traces are available, each query includes its arrival time, input tokens, and generated tokens. Otherwise, InFactPlanner generates per-query traces from aggregate rates using sampled or synthetic token distributions, defining the simulation window and workload intensity.
The \textit{Simulation Engine} then assigns queries to DCs and servers according to the selected load-balancing policy. The default policy is round-robin, while the system allows users to implement custom strategies, such as latency-aware or carbon-aware routing.
Then, for each DC server, the simulator estimates query start times, completion times, active concurrency, and latency based on the throughput, first-token latency, and concurrency limit. The output of this stage is a set of per-query and per-server execution statistics.

The \textit{Sustainability Modeling Stage} enriches these execution statistics with energy, carbon, and water estimates. Server-level power demand is aggregated at the DC level and scaled using its Power Usage Effectiveness~(PUE), which represents how much overhead energy is used to support computing. If RESs are configured, InFactPlanner estimates local generation using either analytical or ML-based models and subtracts it from the total demand to compute the residual grid energy. This residual demand is then aligned with location-specific carbon-intensity data to estimate emissions. Water consumption is estimated using configured Water Usage Effectiveness (WUE) values.

Finally, \textit{Report Builder} combines simulation and sustainability outputs into structured Pandas Dataframes. These outputs offer unified performance and environmental views across queries, servers, DCs, and the overall deployment, supporting comparison through built-in visualizations or external tools.

\vspace{-.1\baselineskip}
\section{Implementation details}
\vspace{-.1\baselineskip}

This section presents the InFactPlanner's implementation details. Major modeling components, including latency modeling, power accounting, RES generation, and reporting, are implemented behind replaceable interfaces. This allows users to retain trace processing and reporting while replacing individual models with alternative or higher-fidelity implementations.

\vspace{-.2\baselineskip}
\subsection{Deployment Configuration \& Input Traces}
\label{sec:configs}
\vspace{-.2\baselineskip}

InFactPlanner is configured through a declarative YAML file, shown in Fig.~\ref{fig:framework} (Configurations). 
The configuration captures the experimental setup, including the simulation start time, input workload, and DC properties.
The \texttt{constant\_TPS} and \texttt{constant\_TTFT} fields select whether throughput and first-token latency are treated as fixed profile values or computed dynamically based on active concurrency and query characteristics.
The \texttt{deployments} field describes one or more DCs. 
Each DC specifies its identifier, geographic location, efficiency parameters ($PUE$, $WUE\_offsite$, $WUE\_onsite$), server groups, and optional RES sources. Each server group defines the hardware type, deployed model, maximum concurrency, and replication factor, allowing a DC to include heterogeneous hardware-model pairings. A hardware-model pairing is characterized by specific parameters, which 
are used by the Simulation Engine and include idle and active power, throughput measured in Tokens Per Second ($TPS$), and median first-token latency. These profiles can be selected from a built-in database or provided by the user.
InFactPlanner intentionally abstracts away low-level serving mechanisms such as batching, KV-cache behavior, and prefill/decode phase separation. The framework targets infrastructure-level planning rather than micro-level execution, and captures serving behavior through 
hardware-model profiles that summarize empirically observed performance in terms of throughput, time to first token, and consumed power. This abstraction enables consistent comparison across deployment configurations, while avoiding the complexity of reproducing every detail of a production serving stack.
Each RES is defined through \texttt{res\_type}, \texttt{model\_type}, and the required parameters, such as \texttt{model\_path} for ML-based generation of solar power production  or \texttt{max\_power} and \texttt{max\_irradiation} for analytical-based estimation.

In addition to the deployment description, users provide an input trace that defines the LLM inference workload to be repeated. The system reads this workload from a CSV file, 
and aligns it with the evaluation start time defined by \texttt{start\_point\_of\_simulation}.
InFactPlanner currently supports two trace formats. The first is a detailed per-query trace, where each request is described by its timestamp, number of context tokens, and number of generated tokens. This trace is provided as a CSV file with columns \texttt{TIMESTAMP}, \texttt{ContextTokens}, and \texttt{GeneratedTokens}. In this case, the workload can be replayed directly by the Simulation Engine.
The second format is an aggregate arrival-rate trace, where users provide the number of incoming queries per second in a CSV file with columns \texttt{TIMESTAMP} and \texttt{InputQueries}. Since this format does not include token-level details, we synthesize a per-query trace before execution. This can be done either by sampling token counts from the Azure Inference Dataset~\cite{azurellminference2024} or by generating synthetic requests with Alibaba's ServeGen tool~\cite{xiang2025servegen}. In both cases, InFactPlanner reconstructs the context-token and generated-token columns so the final trace follows the per-query format.
The per-query trace is passed to the Simulation Engine, where requests are replayed to estimate timing, utilization, and power.

\vspace*{-.1\baselineskip}
\subsection{Trace-Driven Execution Simulation}
\vspace*{-.1\baselineskip}

Then, the Simulation Engine replays the generated workload across the configured DCs and server groups. The engine performs deployment-level execution modeling rather than low-level GPU emulation. It estimates request timing using the configured hardware-model profiles, load-balancing policy, throughput, first-token latency, and concurrency limits.
Specifically, the simulation proceeds in two levels. First, the global query trace is distributed across DCs using the \textit{Inter Load Balancer} and its round-robin policy by default. Then, each DC runs its \textit{Intra Load Balancer} that assigns incoming requests to the available LLM-server instances. The default Intra Load Balancer also follows a round-robin strategy, which provides a simple and reproducible baseline for distributing load across servers. 
Users can implement alternative policies, such as latency-aware, utilization-aware, or carbon-aware routing, by extending the InFactPlanner's \textit{Load Balancer} interface.

For each server instance, the simulator maintains the set of active queries and the available concurrency slots. When a query arrives, it starts immediately if a slot is available. Otherwise, it is queued until the earliest active query completes and releases a slot. The simulator then records the query arrival time, start time, completion time, latency, and active concurrency. Thus, InFactPlanner is able to capture the effect of finite server capacity on request waiting time and utilization.

Query execution can follow either a static or dynamic performance model, depending on the configuration parameters \texttt{constant\_TPS} and \texttt{constant\_TTFT}. When \texttt{constant\_TPS} is set to \texttt{True}, the simulator uses a fixed throughput value for each hardware-model pair and estimates the execution duration of each query using Eq.~\ref{eq:duration}~\cite{jegham2025}.

    \vspace*{-.3\baselineskip}
\begin{equation}
\text{duration} = \text{TTFT} + \frac{\text{generated tokens}}{\text{TPS}}
\label{eq:duration}
\end{equation}
\vspace*{-.8\baselineskip}

In this expression, \textit{generated tokens} denotes the number of output tokens produced by the query, $TPS$ denotes the token-generation throughput, and $TTFT$ denotes the time-to-first-token latency. When \texttt{constant\_TTFT} is set to \texttt{True}, $TTFT$ is taken directly from the hardware--model profile. Otherwise, it is computed dynamically as a function of query characteristics, such as the number of input tokens. Both $TPS$ and $TTFT$ are specific to the selected GPU and LLM combination and can be obtained from the configuration file or from built-in profiles derived from external sources~\cite{artificialanalysis2025}.
When \texttt{constant\_TPS} is set to \texttt{False}, the simulator uses a dynamic throughput profile. In this mode, $TPS$ varies over time according to concurrently active queries per second. The simulator updates the effective throughput as concurrency changes and derives each query’s completion time accordingly.

To reduce execution time, InFactPlanner parallelizes simulation with Apache Ray. Each DC runs as an independent Ray task, while server-level execution within each DC is also parallelized using the assigned trace subsets and hardware-model profiles. This design supports both single-machine and distributed Ray deployments. After all Ray tasks complete, their execution records are collected and passed to the \textit{Report Builder}, which computes latency, power demand, energy consumption, carbon emissions, and water consumption.

\subsection{Renewable Energy Modeling}
\label{sec:res}

After generating the execution statistics, InFactPlanner estimates the renewable energy production associated with each configured DC. This functionality is implemented through an extensible \textit{RES Model} component, invoked by the \textit{Report Builder}. Each RES entry specifies the renewable source type and the selected modeling approach. The current implementation focuses on solar energy and supports two options, namely math-based and ML-based models, while the same interface can be extended to other sources such as wind.
Each RES model uses the DC geolocation to retrieve historical weather data for the simulation period through an external API such as Open-Meteo. For math-based solar generation, the model scales observed solar radiation according to the configured PV capacity. The scaling factor is computed from the maximum PV output and the corresponding maximum irradiation value. For ML-based generation, InFactPlanner loads a pre-trained regressor, currently a \textit{PyCaret} model trained in~\cite{Symeonides2024}. The renewable generation time series is passed to reporting to separate local supply from residual grid demand.

\vspace*{-.1\baselineskip}
\subsection{Energy, Carbon, and Water Accounting}
\vspace*{-.1\baselineskip}

Next, InFactPlanner derives sustainability metrics, including energy consumption, carbon emissions, and water use, by enriching the simulator's per-query and per-server records with power and environmental estimates. 
All accounting components are modular, allowing users to update configurations, as described in Sec.~\ref{sec:configs}, or replace individual models.

First, InFactPlanner estimates the power demand associated with query execution. For each server and time step $t$ (default value is $1sec$), the instantaneous power of the server, denoted by $P_{\mathrm{node}}(t)$, is modeled as $P_{\mathrm{node}}(t) = x_t \left(P_{\mathrm{max}} - P_{\mathrm{idle}}\right) +P_{\mathrm{idle}}$, where $P_{\mathrm{idle}}$ and $P_{\mathrm{max}}$ denote the minimum and maximum node power draw, respectively. The variable $x_t$ is a sampled load-dependent factor representing instantaneous node workload intensity, allowing power consumption to scale between idle and peak power. Following~\cite{oviedo2024energy}, we sample $x_t$ from a truncated log-normal distribution parameterized using the reported P5--P95 range.
To estimate query-level power, the sampled node power is amortized across concurrent requests:

\vspace*{-.3\baselineskip}
\begin{equation}
P_{\mathrm{query}}(t) = \frac{P_{\mathrm{node}}(t)}{n(t)},
\end{equation}
\vspace*{-.5\baselineskip}

\noindent where $n(t)$ is the number of concurrent requests at time~$t$. 
Thus, $P_{\mathrm{query}}$ and $P_{\mathrm{node}}(t)$ are accounting estimates, replaceable by measured traces or power distributions when available.

The corresponding facility-level server power is computed by applying the DC's PUE, defined as the ratio of total facility energy to IT energy:
$P_{\mathrm{server}}(t) = P_{\mathrm{node}}(t) \cdot \mathrm{PUE}$.
Server-level energy over a time interval $D$ is then obtained by integrating facility-level power over time. Since InFactPlanner operates at one-second granularity, this integration is implemented as a discrete sum and converted to watt-hours:

\vspace*{-.5\baselineskip}
\begin{equation}
E_{D} = \frac{1}{3600}\sum_{t=0}^{D-1} P_{\mathrm{server}}(t).
\label{eq:energy}
\end{equation}
\vspace*{-.5\baselineskip}

\noindent where power is expressed in watts and each time step corresponds to one second.
Per-query energy is estimated by multiplying each query's simulated execution duration by its average power and the DC's PUE: 
\begin{equation}
E_{\mathrm{query}}~=~\left(\mathrm{Duration}/3600\right)~\cdot~\overline{P}_{\mathrm{query}}\cdot~\mathrm{PUE}
\label{eq:query_energy}
\end{equation}
where $\overline{P}_{\mathrm{query}}$ is the average amortized query power over the query's execution interval.
This duration-based estimate assumes the same average power draw during prompt processing and output-token generation. This abstraction matches InFactPlanner's deployment-level scope and can be replaced by phase-specific power models when available.

Next, InFactPlanner aggregates the facility-level power demand of all servers to obtain the total DC power demand at each time step. It then invokes the RES models described in Sec.~\ref{sec:res} to compute the power generated by each local renewable source. The total RES production of a DC~($d$), denoted as $P_{\mathrm{RES},d}(t)$, is subtracted from the DC demand to estimate the residual power drawn from the electricity grid: $P_{\mathrm{grid},d}(t) = \max\left(0, P_{\mathrm{DC},d}(t) - P_{\mathrm{RES},d}(t)\right)$.
The corresponding grid energy, $E_{\mathrm{grid}}$, is computed using the same discrete integration process as in Eq.~\ref{eq:energy}. InFactPlanner then estimates carbon emissions by multiplying the time-step grid energy, expressed in kWh, by the time-aligned carbon intensity of the local electricity grid: $\mathrm{Emissions}(t) = E_{\mathrm{grid}}(t)\cdot \mathrm{CI}(t)$.
Carbon intensity, denoted as $\mathrm{CI}$, is expressed in grams of CO$_2$ equivalent per kilowatt-hour (gCO$_2$eq/kWh), and is computed as a weighted average of source-specific carbon~factors:

\vspace*{-.3\baselineskip}
\begin{equation}
\mathrm{CI} = \frac{\sum_i E_i f_i}{\sum_i E_i}
\label{eq:carbon_intensity}
\end{equation}
\vspace*{-.5\baselineskip}

\noindent where $E_i$ is the electricity generated by source $i$ and $f_i$ is the corresponding carbon factor. 
Per-source electricity generation and historical carbon factors are sourced from ENTSO-E and ElectricityMaps~\footnote{\url{https://app.electricitymaps.com/map} \& \url{https://www.entsoe.eu/}}, respectively. Since carbon intensity varies over time with the grid mix, InFactPlanner aligns the residual grid demand with the carbon-intensity time series. We assume that electricity generated by on-site RES has zero operational carbon emissions. If RES generation exceeds the DC demand at a given time step, the residual grid demand is set to zero.

Finally, InFactPlanner estimates water needs using on-site and off-site WUE coefficients. On-site WUE captures water drawn directly at the DC, for example for cooling, and is applied to IT energy. Off-site WUE captures water associated with electricity generation and is applied to facility-level electrical energy. Water use is computed as:

\vspace*{-.7\baselineskip}
\begin{equation}
\mathrm{Water} = \frac{E_{\mathrm{DC}}}{\mathrm{PUE}}\cdot \mathrm{WUE}_{\mathrm{onsite}} + E_{\mathrm{DC}}\cdot \mathrm{WUE}_{\mathrm{offsite}}
\label{eq:water}
\end{equation}
\vspace*{-.7\baselineskip}

\noindent where $E_{\mathrm{DC}}$ is the facility-level energy consumed during the interval. Since $\mathrm{PUE}$ relates facility energy to IT energy, $\frac{E_{\mathrm{DC}}}{\mathrm{PUE}}$ gives the IT-energy basis used for on-site WUE. Both WUE coefficients are expressed in liters per kWh~\cite{jegham2025}.

\section{Framework Validation and What-if Analysis}
\label{sec:eval}
We evaluate InFactPlanner as an early-stage deployment-level model, focusing on accounting consistency, scalability, and comparative deployment trade-offs rather than low-level LLM serving behavior. First, we assess whether it reproduces reference per-query energy and latency estimates under aligned assumptions. Second, we measure wall-clock simulation time as deployment size increases. Third, we use controlled what-if scenarios to quantify how hardware-model selection, RES placement, site carbon intensity, and workload routing affect latency, energy, carbon emissions, and water consumption.
All trials were run on an HP ProLiant DL380 Gen9 server with 48 logical CPU cores, 192 GB RAM, and Ubuntu 22.04 LTS. 

\vspace{-0.1\baselineskip}
\subsection{Real-World Production-Grounded Validation}
\vspace{-0.1\baselineskip}

\begin{table}[t]
\centering
\caption{Energy validation results for DeepSeek-R1 on DGXH100}
\vspace{-0.5\baselineskip}
\label{tab:validation}
\scriptsize
\setlength{\tabcolsep}{3pt}
\renewcommand{\arraystretch}{1.1}
\begin{tabular}{llccc}
\hline
\textbf{Workload} & \textbf{Metric} & \textbf{InFactPlanner} & \textbf{Ref. study} & \textbf{Err.} \\
\hline
Traditional
& Median & 0.65 & 0.70 & 7.18\% \\
& IQR Low / IQR High & 0.27 / 1.34 & 0.28 / 1.42 & 4.06\% / 5.37\% \\
\hline
Reasoning
& Median & 10.66 & 11.68 & 8.71\% \\
& IQR Low / IQR High & 4.45 / 21.80 & 4.73 / 23.91 & 5.86\% / 8.82\% \\
\hline
\end{tabular}
\vspace{-2\baselineskip}
\end{table}

We first verify that InFactPlanner reproduces the energy and latency estimates reported in two prior real-system studies. 
The first study~\cite{oviedo2024energy} derives its estimates from energy metrics collected on a production-grade DGX H100 system executing real LLM inference workloads, for which we focus on its DeepSeek-R1 configuration. The second study~\cite{mlenergy} provides energy and latency traces from real-world deployments, and we use its DeepSeek-R1 deployment on DGX B200, as an H100-based setup is not available. Through these trials, we evaluate whether InFactPlanner can realistically reproduce energy and latency distributions for real-world LLM inference deployments under aligned workload and power-model assumptions, while intentionally abstracting fine-grained serving behaviors such as batching and KV-cache dynamics.

To ensure consistency, we align InFactPlanner’s parameters with the assumptions of the reference studies. For the DeepSeek-R1 on DGX H100 setup~\cite{oviedo2024energy}, we adopt the same maximum power ($P_{\max}$), model node power using the same log-normal distribution, and set the $PUE$ to the median reported value. For throughput, we use the same formulation as the reference study and evaluate Tokens Per Second (TPS) at the median query characteristics of each workload. To match the study’s simplified formulation, we set the time-to-first-token latency (TTFT) to zero, so that query duration depends solely on token generation time, while concurrency is fixed at one. We then synthesize two traces of 10k queries matching the reported workloads: (i) traditional queries with 500 input tokens and short outputs sampled from an exponential distribution with median 300 tokens and IQR 129--618, and (ii) reasoning queries with 500 input tokens and longer outputs with median 5k tokens and IQR 2,040--9,717. For the second study~\cite{mlenergy}, we similarly adopt the study's $TPS$, $TTFT$, and $concurrency$ parameters for the DeepSeek-R1 on DGX B200 setup, and replay the study's reported prompt traces.

\begin{figure}[!t]
    \centering
    \includegraphics[width=0.99\linewidth,
     trim={0.3cm 0.3cm .1cm 0.2cm},
    clip]{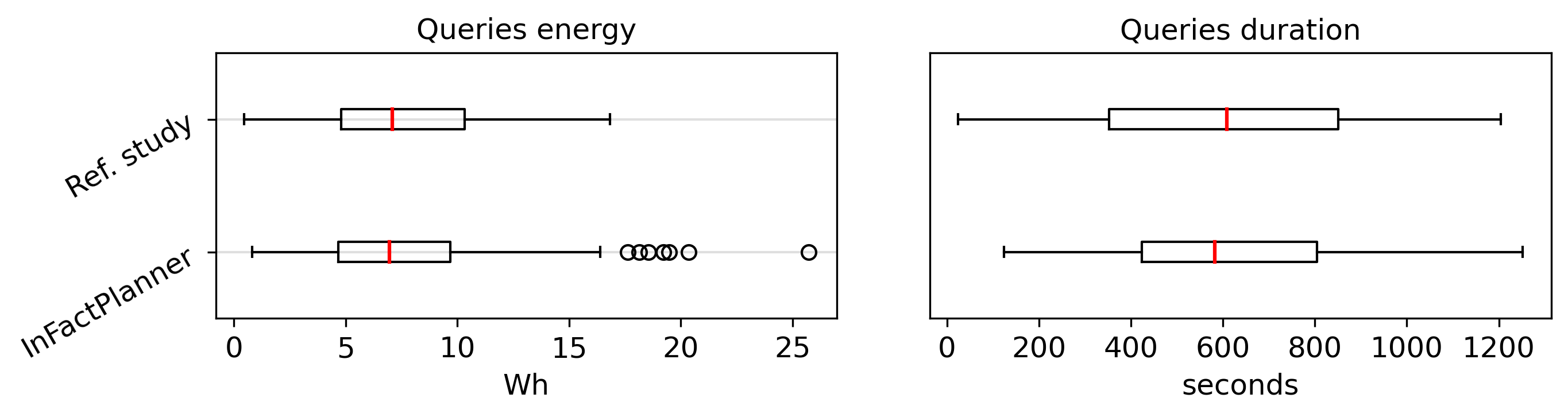}
    \vspace*{-1.5\baselineskip}
    \caption{Energy and Latency validation results for DeepSeek-R1 on DGXB200.}
    \label{fig:boxplots_deepseek}
    \vspace*{-1.3\baselineskip}
\end{figure}

Table~\ref{tab:validation} compares InFactPlanner with the reference results from~\cite{oviedo2024energy} using median per-query energy, IQR, and percentage error. InFactPlanner reproduces the reference accounting estimates with deviations below 9\%. This holds for both median and IQR values, 
preserving the energy-distribution characteristics observed on the measured DGX H100 deployment. It also captures the substantial increase in energy consumption caused by reasoning workloads, a key finding of the reference study.
Fig.~\ref{fig:boxplots_deepseek} further compares the real and simulated energy and latency distributions of the second study~\cite{mlenergy}, providing a more detailed view of their variability.
The median differences in energy and prompt duration across all queries are 1.9\% and 4.5\%, respectively, while the corresponding IQRs remain closely aligned. Minor deviations appear in the lower query duration range, likely due to $TTFT$ variations, while the simulated energy distribution includes a few higher-value samples absent from the reference measurements. Overall, \textit{InFactPlanner reproduces reference energy and latency trends with low deviation, supporting its use for comparative deployment planning.}

\vspace{-0.3\baselineskip}
\subsection{Scalability Evaluation}
\vspace{-0.3\baselineskip}
In this section, we study how InFactPlanner scales with problem size by reporting the wall-clock time. 
The input trace captures one day of queries from the Azure LLM Inference Dataset~\cite{azurellminference2024}. 
The evaluated traces contain 1.24M, 2.48M, 4.96M, and 9.92M queries for the 10, 20, 40, and 80-server one-day configurations, respectively. 
First, we fix the number of servers per DC to 10 and vary the number of DCs across 1, 2, 4, and 8. 
The process completes in 4, 5, 6, and 9 minutes, respectively, showing the benefit of parallelization across DCs and servers, since runtime does not grow proportionally with the number of queries and DCs.
Then, we consider a single DC and vary the number of servers per site across 10, 20, 40, and 80, matching the total number of servers in the previous experiment. 
In this case, runtime increases from 4 to 7, 14, and 28 minutes, respectively, showing a more proportional increase because parallelization occurs only across servers within one DC.
We should note that, for the one-day trace, InFactPlanner completed in under 30 minutes across all configurations, while more detailed simulators~\cite{cho2026llmservingsim} report over 10 minutes for traces shorter than 3 minutes. Moreover, across our experiments, InFactPlanner used at most 25GB of memory, showing that it can run on commodity servers without specialized hardware.
These results indicate that \textit{the abstraction level adopted by InFactPlanner enables tractable exploration of large deployment design spaces using commodity infrastructure}.


\vspace{-0.2\baselineskip}
\subsection{Scenario-based What-if Case Studies}
\vspace{-0.2\baselineskip}


We evaluate four what-if scenarios over the main deployment decisions exposed by InFactPlanner, namely hardware--model selection, renewable integration, multi-DC carbon intensity, and a scenario with carbon-aware load balancing policies. Each scenario is specified through a YAML configuration and runs end-to-end without code changes, allowing us to isolate how individual configuration choices affect latency, energy, carbon emissions, and water use. We use performance metrics from~\cite{artificialanalysis2025} and GPU power requirements from NVIDIA's documentation. 
We sample the Azure LLM Inference Dataset~\cite{azurellminference2024} to obtain a one-week trace corresponding to a 10-server DC.
We also transferred the trace to start from October 5, 2025, and we kept the days of week the same as in the original dataset, with the DC geolocation being near Athens, Greece~($GR$).
Rather than focusing on a single fixed deployment or an explicit parameter-sensitivity analysis, these scenarios assess how predicted sustainability and performance outcomes vary across alternative infrastructure configurations.



\begin{figure}[!t]
    \centering
    \includegraphics[width=0.95\linewidth,
    trim={0.4cm 0.4cm .3cm 0.22cm},
    clip]{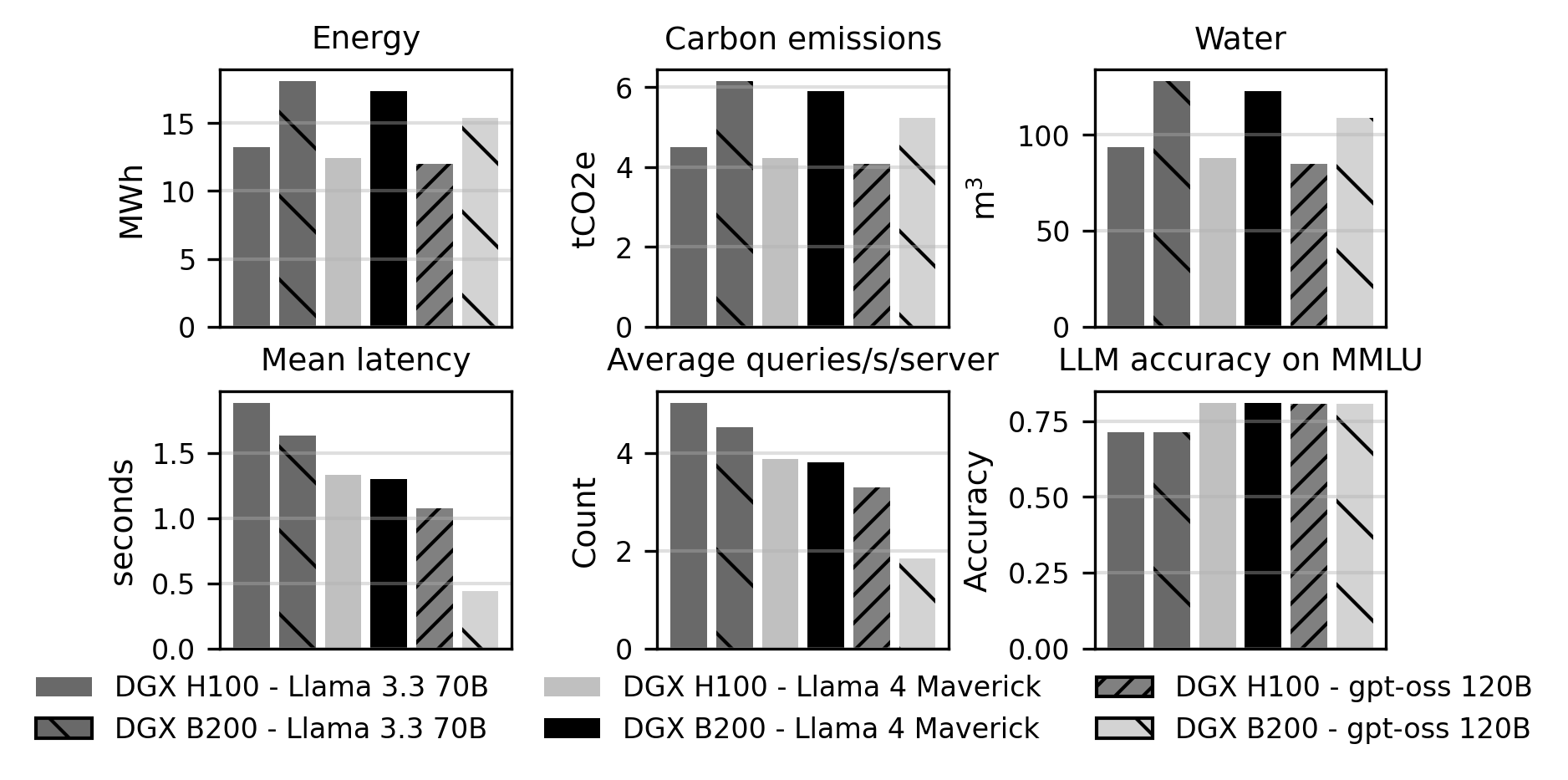}
    \vspace*{-.7\baselineskip}
    \caption{Metrics for different GPU-LLM combinations.}
    \label{fig:all}
    \vspace*{-1.7\baselineskip}
\end{figure}

\subsubsection{Performance of LLM and Hardware Pairings}\label{sec:different}

We first compare six hardware--model pairings using DGX H100 and DGX B200 servers with Llama 3.3 70B, Llama 4 Maverick, and gpt-oss 120B, while keeping the site, trace, and efficiency parameters fixed.
Fig.~\ref{fig:all} reports the resulting metrics.
The Llama models consume up to 50\% more energy and produce up to 50\% higher carbon emissions, mainly because of higher first-token latency and lower $TPS$ than gpt-oss, as reported in~\cite{artificialanalysis2025}.
DGX B200 pairings further increase the environmental footprint by 35\% due to higher system power, but reduce mean latency by up to 77\% and active queries per server by 63\%.
Despite these performance gains, the footprint increases because the one-week query trace contains substantial idle periods, during which DGX B200 GPUs draw more idle power than DGX H100 GPUs.
Since Llama 4 and gpt-oss 120B also provide higher accuracy based on the MMLU benchmark~\cite{hendrycks2021mmlu, artificialanalysis2025}, the \textit{preferred deployment depends on whether the operator prioritizes environmental impact, latency, or accuracy.}
For the remaining scenarios, we use \textit{10 DGX H100 nodes with gpt-oss 120B} as the default plan.

\subsubsection{Single-site deployments with RES integration}\label{sec:first}
\begin{figure}[!t]
    \centering
    \includegraphics[width=0.85\linewidth,
    trim={0.2cm 0.2cm .25cm 0.27cm},
    clip]{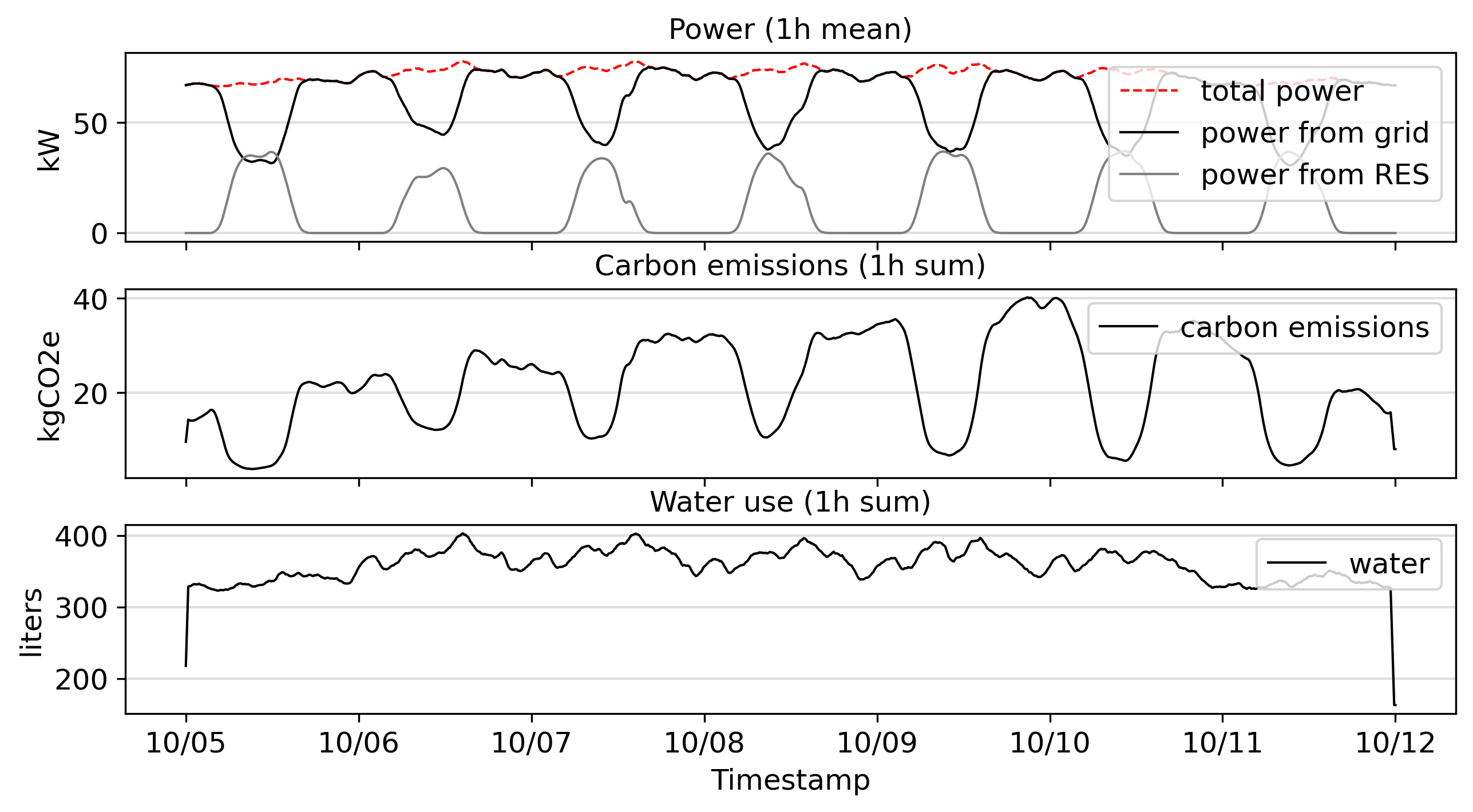}
    \vspace*{-.7\baselineskip}
    \caption{DC with 10 DGX H100 GPUs - gpt-oss 120B and local RES}
    \label{fig:exp1}
    \vspace*{-.5\baselineskip}
\end{figure}

Next, we keep the baseline deployment and attach on-site RES using ML-based modules from~\cite{Symeonides2024}. This experiment verifies that users can model local generation and obtain aligned time series for power demand, residual grid draw, carbon emissions, and water consumption. Fig.~\ref{fig:exp1} reports the resulting power, carbon, and water trends over one week. The first plot shows the average power on an hourly basis in three forms: the red dashed line shows the total power required by the DC, the grey line shows the power supplied by local RES, and the black line indicates the power drawn from the grid. The remaining plots show the corresponding carbon emissions and water consumption of that period.
Thus, \textit{the estimated footprint is shaped by the temporal alignment between workload demand, local RES generation, and residual grid dependence.}

\subsubsection{Multi-site DCs under Heterogeneous Grids and RES}\label{sec:2countries}
\begin{figure}[!t]
    \centering
    \includegraphics[width=0.85\linewidth,
    trim={0.2cm 0.22cm .25cm 0.27cm},
    clip]{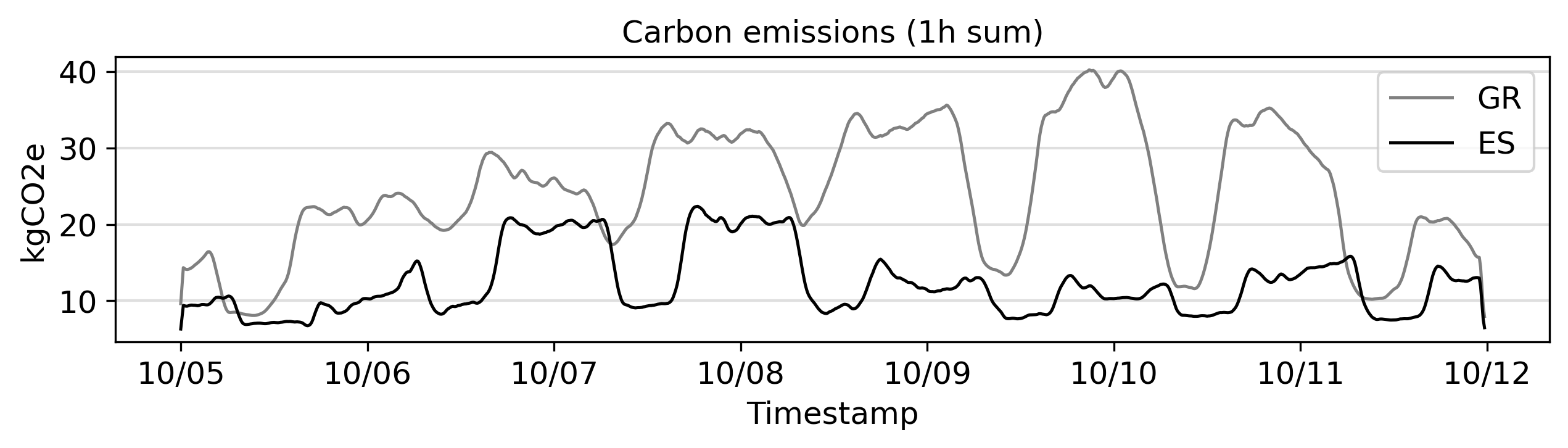}
    \vspace*{-1.\baselineskip}
    \caption{Timeline of carbon emissions for two countries.}
    \label{fig:exp3}
    \vspace*{-.5\baselineskip}
\end{figure}

Next, we consider an operator who plans to build two DCs in Greece~($GR$) and Spain~($ES$) by extending the baseline to two sites, with the same efficiency metrics ($PUE$, $WUE$), but with different carbon-intensity time-series, and without RES. 
The two DCs share the same input trace, so we can compare their environmental impact. Fig.~\ref{fig:exp3} presents the time series of carbon emissions for the two countries. $GR$ emits more $CO2$ than $ES$, due to differences in the electricity generation mix.

\begin{figure}[!t]
    \centering
    \includegraphics[width=0.9\linewidth,
    trim={0.2cm 0.22cm .25cm 0.26cm},
    clip]{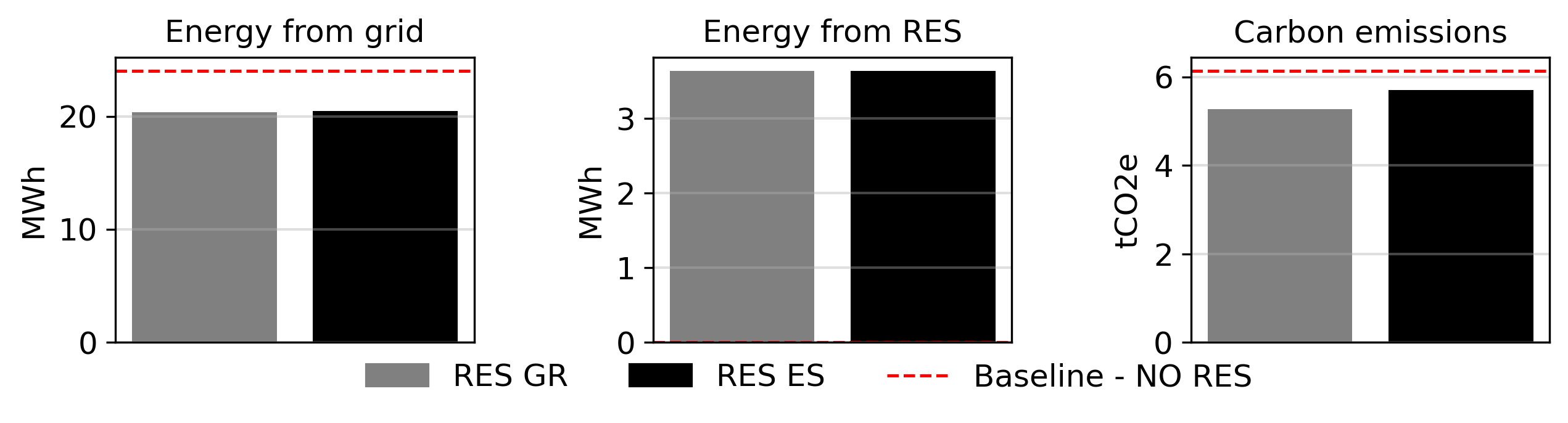}
    \vspace*{-1.\baselineskip}
    \caption{Total metrics for two deployments with renewable integration.}
    \label{fig:exp3_barplot}
    \vspace*{-1.5\baselineskip}
\end{figure}
The operator then aims to establish a PV panel park in one of the two countries and must determine the optimal location. The previous configuration is extended by incorporating RES for $GR$ and subsequently for $ES$, allowing us to calculate the total environmental impact associated with energy needs and carbon emissions. 
Specifically, Fig.~\ref{fig:exp3_barplot} reports the total grid energy, RES energy, and carbon emissions for the two deployments ($RES\ GR$, $RES\ ES$), alongside the baseline without RES (red dashed line). The grid and RES energy values of the two deployments differ by only 0.3\%, owing to their relatively similar latitudes and timezones, which result in comparable solar energy production patterns throughout the day. In contrast, placing RES in $GR$ reduces carbon emissions by 7.5\% compared to $RES\ ES$, because it displaces electricity from a more carbon-intensive grid.
This scenario shows that \textit{RES placement depends not only on RES output, but also on the carbon intensity of the displaced grid energy.}

\subsubsection{Carbon-Aware Utilization-Balanced Routing}
Next, the operator evaluates carbon and utilization-aware routing across sites. For each incoming request, the Inter Load Balancer computes a score for each candidate site using its effective carbon cost and utilization. The effective carbon cost is defined as $E_i=\max(0, C_i-w_rR_i)$, where $C_i$ is the normalized grid carbon intensity, $R_i$ the normalized RES availability, and $w_r$ controls how much RES mitigates the carbon penalty.
The final score is $S_i=w_cE_i+w_uU_i$, where $w_c$ sets the importance of carbon-aware placement, $w_u$ controls the penalty for already utilized sites, and $U_i$ is estimated from a running count of previous assignments. Each request is routed to the site with the lowest score, favoring low-carbon and RES-rich periods while avoiding persistent workload concentration at one site.

\begin{figure}[!t]
    \centering
    \includegraphics[width=0.95\linewidth,
    trim={0.2cm 0.22cm .25cm 0.265cm},
    clip]{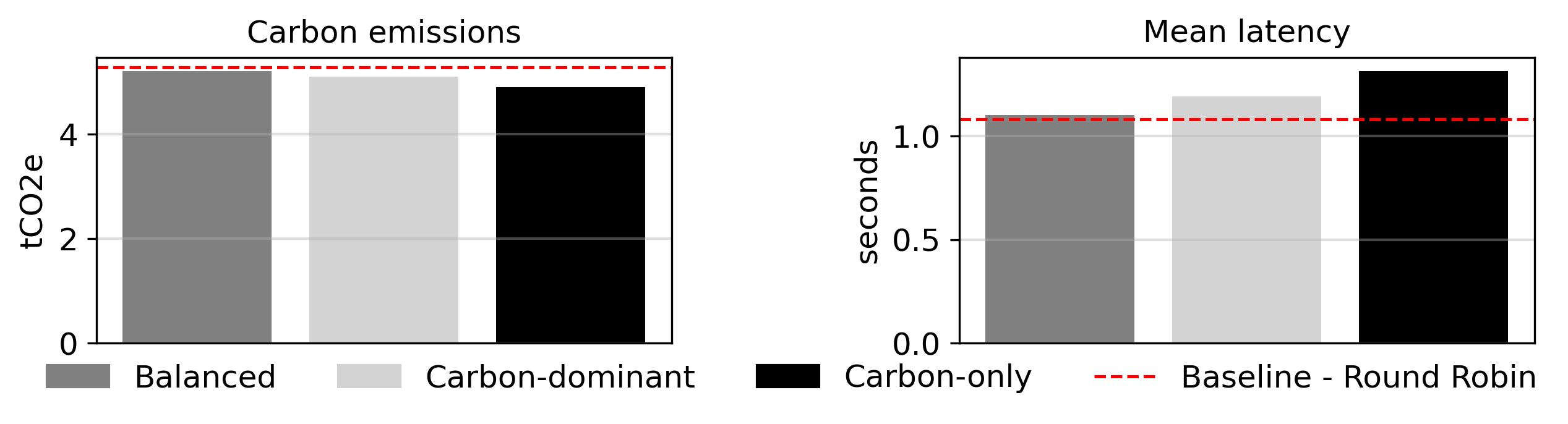}
    \vspace*{-.7\baselineskip}
    \caption{Metrics for different load balancing algorithms.}
    \label{fig:distributions}
    \vspace*{-1.5\baselineskip}
\end{figure}

We compare three illustrative prompt routing policies against the default round-robin policy: carbon-only $(w_c=1.0,w_u=0.0)$, balanced $(w_c=0.5,w_u=0.5)$, and carbon-dominant $(w_c=0.75,w_u=0.25)$. In all cases, $w_r=1.0$. We omit utilization-only routing since, when sites have equal capacity, it behaves similarly to round-robin.
Fig.~\ref{fig:distributions} (left) shows the carbon impact of the routing policies. Carbon-only routing achieves the largest carbon reduction, lowering emissions by 7\% compared to round-robin, but it also concentrates more requests at the cleaner site. Carbon-dominant routing provides a slightly smaller reduction of 3\% while reducing this imbalance, whereas the balanced policy achieves a more moderate reduction of 1\% and keeps assignments closer to round-robin.
Moreover, Fig.~\ref{fig:distributions} (right) shows the performance impact, with carbon-only routing increasing latency by 22\%. The balanced policy keeps latency within 2\% of round-robin, while carbon-dominant routing offers an intermediate trade-off. Overall, \textit{carbon-aware routing reduces emissions without changing the infrastructure, but the routing weights ($w_r$,~$w_c$,~$w_u$) determine the carbon--performance trade-off.}

    \vspace*{-.2\baselineskip}
\subsection{Validity Considerations}
    \vspace*{-.2\baselineskip}


InFactPlanner targets deployment-level what-if analysis rather than low-level GPU serving simulation. It abstracts fine-grained effects such as batching, KV-cache pressure, memory contention, and prefill/decode separation through configurable hardware--model profiles. This improves scalability and enables rapid comparison across deployment options, but may deviate from production serving stacks under highly dynamic workloads.
The accuracy of its environmental accounting depends on the fidelity of input models and datasets, including hardware power profiles, throughput estimates, RES models, weather traces, grid carbon intensity, and WUE parameters. Since the evaluated scenarios depend on selected traces, assumptions, and routing configurations, the results should be interpreted as comparative planning studies rather than exact production-footprint certification. Finally, InFactPlanner excludes embodied carbon and does not credit surplus RES exports; excess generation only reduces grid draw to zero.


\vspace*{-.2\baselineskip}
\section{Conclusion}
\vspace*{-.2\baselineskip}

In this paper we introduced a trace-driven analytical framework for estimating the sustainability footprint of LLM inference across geo-distributed DCs. Our evaluation shows that InFactPlanner can reproduce real-system energy-accounting trends, scale to multi-million-query workloads, and support deployment-level what-if analysis across hardware selection, renewable integration, placement, and routing decisions.
In \textit{future work}, we plan to validate InFactPlanner with operational DC traces, measured power behavior, and production constraints to refine its accounting models under realistic conditions. We also plan to extend it beyond centralized DCs toward cloud-edge deployments, where LLM inference may span large DCs, edge sites, and constrained edge devices. This will enable the study of placement, routing, and sustainability trade-offs across heterogeneous infrastructures with different latency, energy, carbon, and resource characteristics.

\bibliographystyle{ieeetr}
\bibliography{references}    
\end{document}